\documentclass[12pt,preprint]{aastex}
\usepackage{graphics}
\usepackage{lscape, graphicx}
\usepackage{apjfonts}
\usepackage{natbib}

\newcommand{\etal}{et al.}  
\newcommand{\per}{\ensuremath{^{-1}}}

\newcommand{\hal}{H\ensuremath{\alpha}}

\newcommand{\msun}{\ensuremath{M_{\odot}}}
\newcommand{\kms}{km~s\ensuremath{^{-1}}} 
\newcommand{\ergs}{erg~s\ensuremath{^{-1}}}

\newcommand{\mbh}{\ensuremath{M_\mathrm{BH}}}

\newcommand{\sigmastar}{\ensuremath{\sigma_{*}}}

\newcommand{\msigma}{\ensuremath{\mbh-\sigma_{*}}}

\newcommand{\rosat}{\emph{ROSAT}}
\newcommand{\spitzer}{\emph{Spitzer}}

\newcommand{\xmmn}{\emph{XMM-Newton}}
\newcommand{\chandra}{\emph{Chandra}}
\newcommand{\oiii}{\ensuremath{\mathrm{[O~III]}}}

\newcommand{\lbol}{\ensuremath{L_{\mathrm{bol}}}}

\newcommand{\ledd}{\ensuremath{L_{\mathrm{Edd}}}}
\newcommand{\loiii}{\ensuremath{L_{\oiii}}}
\newcommand{\lX}{\ensuremath{L_{\mathrm{X}}}}

\newcommand{\one}{SDSS~J011905}
\newcommand{\ten}{SDSS~J103234}
\newcommand{\eleven}{SDSS~J110912}
\newcommand{\fourteen}{SDSS~J144012}

\slugcomment{{\it Accepted for publication in \apj}}

\shorttitle{XMM-NEWTON OBSERVATIONS OF LOW-MASS  SEYFERT 2 GALAXIES} 
\shortauthors{THORNTON ET AL.}

\begin{document}
\title{Emission and Absorption Properties  of Low-Mass Type 2 Active Galaxies with {\it XMM-NEWTON}\altaffilmark{1}} 
\author {Carol E. Thornton\altaffilmark{2}, Aaron J. Barth\altaffilmark{2}, Luis C. 
Ho\altaffilmark{3}, and Jenny E. Greene\altaffilmark{4,5}.}

\altaffiltext{1}{Based on observations obtained with XMM-Newton, an ESA science mission with 
instruments and contributions directly funded by ESA Member States and NASA.}

\altaffiltext{2}{Department of Physics \& Astronomy, 4129 Frederick Reines Hall, University of 
California, Irvine, Irvine, CA 92619-4575; thorntoc@uci.edu}
\altaffiltext{3}{The Observatories of the Carnegie Institution for Science, 813 Santa Barbara 
Street, Pasadena, CA 91101.}
\altaffiltext{4}{Department of Astrophysical Sciences, Princeton University, Princeton, NJ 
08544}
\altaffiltext{5}{Hubble Fellow and Princeton-Carnegie Fellow}

\begin{abstract}
We present \xmmn\ observations of four low-redshift Seyfert galaxies selected to have low host luminosities ($M_g > -20$ mag) and small stellar velocity dispersions (\sigmastar\ $< 45$ \kms), which are among the smallest stellar velocity dispersions found in any active galaxies. These galaxies show weak or no broad optical emission lines and have likely black hole masses  $\lesssim 10^{6}$~\msun. Three out of four objects were detected with $> 3\sigma$ significance in $\sim 25$ ks exposures and two observations had high enough signal-to-noise ratios for rudimentary spectral analysis. We calculate hardness ratios ($-0.43$ to $0.01$) for the three detected objects and use them to estimate photon indices in the range of $\Gamma = 1.1-1.8$. Relative to [\ion{O}{3}], the type 2 objects are X-ray faint in comparison with Seyfert 1 galaxies, suggesting that the central engines are obscured. We estimate the intrinsic absorption of each object under the assumption that the [\ion{O}{3}] emission line luminosities are correlated with the unabsorbed X-ray luminosity.  The results are consistent with moderate ($N_{\rm H}~\sim~10^{22}~{\rm cm^{-2}}$) absorption over the Galactic values in three of the four objects, which might explain the non-detection of broad-line emission in optical spectra. One object in our sample, SDSS J110912.40+612346.7, is a near identical type 2 counterpart of the late-type Seyfert 1 galaxy NGC~4395. While the two objects have very similar [\ion{O}{3}] luminosities, the type 2 object has an X-ray/[\ion{O}{3}] flux ratio nearly an order of magnitude lower than NGC~4395. The most plausible explanation for this difference is absorption of the primary X-ray continuum of the type 2 object, providing an indication that obscuration-based unified models of active galaxies can apply even at the lowest luminosities seen among Seyfert nuclei, down to \lbol\ $\sim 10^{40}-10^{41}$ \ergs.\end{abstract}


\section{Introduction}
Due to the advent of large area surveys in the past few years, extensive progress has been made in the search for low-mass active galactic nuclei (AGNs) with estimated black hole masses $\mbh \lesssim10^{6}~\msun$. Through searches for galaxies with low stellar velocity dispersions or weak broad-line emission in the Sloan Digital Sky Survey (SDSS), the number of candidate broad-line type 1 \citep{GH04, GH07c} and narrow-line type 2 \citep{BGH08} low-mass AGNs has increased to number in the hundreds. Multi-wavelength studies (e.g. Gallo \etal\ 2008, Satyapal \etal\ 2008) have begun to provide new avenues for finding these low-mass and low-luminosity AGNs that would not typically be identified in the optical, allowing us to observe a larger portion of the total energy output. X-ray observations are of key importance for detecting the primary ionizing continuum of the AGN and determining the total luminosity as well as constraining the obscuration toward the central engine.

Current unification schemes explain the observational differences between type 1 and type 2 AGNs by the viewing angle from which we observe the central engine. For a type 2 object, an obscuring torus blocks the light coming from the innermost region, which contains the broad emission-line and X-ray-emitting regions. Therefore in the classical AGN unification picture \citep{AM85}, a type 2 object will show the underlying properties of a type 1 object if the effects of the obscuration are removed. Hence, type 1 and 2 objects with comparable black hole masses and luminosities should also show similar optical narrow emission-line spectra, since the narrow-line emitting region remains largely unobscured. By systematically searching for galaxies of both types 1 and 2 with similar stellar velocity dispersions and narrow emission-line luminosities, one can develop comparable samples of type 1 and type 2 objects with similar black hole masses to test if the unified model is applicable across the full range of black hole masses found in AGNs, or if the lack of broad emission lines in low-mass type 2 objects results from changes in the structure of AGNs at low bolometric luminosities \citep{Nicastro00, Laor, ES06} rather than due to absorption.

The two best studied AGNs with $\mbh~<~10^{6}~\msun$ are located in the late-type spiral NGC~4395 \citep{FS89} and in the dwarf elliptical POX~52 \citep{Kunth87, Barth04}, respectively. NGC~4395 has been shown to vary rapidly in the X-ray \citep{Iwasawa00, Shih03, Moran05}, including dramatic changes in spectral slope ($\Gamma~\approx~0.6-1.7$, Moran \etal\ 2005) over a few years. POX~52 shows similar rapid variability, along with substantial changes in the absorbing column density in $< 1$~year \citep{Thornton}. Additionally, \xmmn\ and \chandra\ have been used to investigate the X-ray properties of low-mass type 1 AGN samples \citep{GH07a, Desroches, Mini} showing that they seem to be scaled down versions of their more massive counterparts with similar hardness ratios and photon indices. The X-ray properties of the type 2 counterparts of these AGNs have not previously been studied systematically.

We present \xmmn\ observations of four galaxies selected from the low-mass Seyfert 2 sample of \citet{BGH08} to have the lowest stellar velocity dispersions in the sample in order to quantify the absorption and emission properties of this population. Our goal is to investigate whether obscuration can explain the lack of broad-line emission in low-mass Seyfert 2 galaxies. From measurements of X-ray luminosities and spectral fitting, we estimate intrinsic absorbing column densities, bolometric luminosities, and corresponding Eddington ratios in order to investigate the differences between type 1 and type 2 objects in this mass range. Throughout this paper we assume $H_{0}~=~70$~\kms~Mpc\per, $\Omega_m$ = 0.3, and $\Omega_\Lambda = 0.7$. All estimates of Galactic foreground column densities are calculated based on Galactic \ion{H}{1} maps \citep{DL90, Kalberla05}, using the HEASARC online $N_H$ calculator\footnotemark.  

\footnotetext{http://heasarc.gsfc.nasa.gov/cgi-bin/Tools/w3nh/w3nh.pl}

\begin{deluxetable}{lllcc}
\tablecaption{Observation Details}
\tablehead{
\colhead{Galaxy} & \colhead{Obs. ID} & \colhead{Obs. Date} & \colhead{Exp. Time} 
& \colhead{Cor. Exp.} \\
\colhead{} & \colhead{} & \colhead{(UT)} & \colhead{(s)} & \colhead{Time (s)}}
\startdata
SDSS J011905.14+003745.0  & 0400570301	& 2006 Jul 26  & 26,739 & 18,040 \\
SDSS J103234.85+650227.9  & 0400570401	& 2006 May 6   & 24,013 & 19,354 \\
SDSS J110912.40+612346.7  & 0400570201	& 2006 Nov 25  & 23,613 & 23,613 \\
SDSS J144012.70+024743.5  & 0400570101	& 2006 Aug 8   & 22,915 & 17,120 \\
\enddata
\tablecomments{Cor. Exp. Time is the exposure time of each observation, corrected for the time lost due to background flares.}
\label{obsDate}
\end{deluxetable}

\section{Sample Selection}
Barth \etal\ (2008, hereafter BGH08) searched SDSS for nearby low-mass active galaxies with absolute magnitudes fainter than $M_g = -20$ mag and emission-line ratios consistent with a Seyfert 2 classification \citep{Ho, Kauffmann, Kewley}. They obtained stellar velocity dispersions from high-resolution Keck spectra and found $39 < \sigmastar < 95$ \kms\ for the sample of 29 galaxies. These type 2 galaxies have a similar range of [\ion{O}{3}] line luminosities and stellar velocity dispersions as the sample of type 1 objects found by \citet{GH04, GH07c}. We selected the four objects from the BGH08 sample with the lowest stellar velocity dispersions and therefore the lowest estimated black hole masses for X-ray observations.  

{\it SDSS J011905.14+003745.0} ($z~=~0.0327$): This galaxy has the smallest measured stellar velocity dispersion, $\sigmastar~=~39~\pm~8$, in the BGH08 sample. Optical spectropolarimetry data obtained by BGH08 found no polarized emission lines or a polarized continuum. 

{\it SDSS J103234.85+650227.9} ($z~=~0.0056$): Also known as NGC~3259, which is designated as a Hubble type SBbc. This is the nearest galaxy in the BGH08 sample. High-resolution Keck spectra confirm the presence of weak broad \hal\ emission. Soft X-ray emission from this object has been previously detected by \rosat\ \citep{Boller92, Moran96}. \citet{Seth} also note it as an example of a galaxy with both an AGN and a nuclear star cluster. BGH08 did not list a stellar mass for this galaxy, so using the prescription of \citet{Bell03} and the SDSS catalogue Petrosian magnitudes ($u = 14.94, r = 12.88$ and $z=12.49$), we estimate a stellar mass of $\mathrm{log}\ (M_\star/\msun) \approx 9.87$. 

{\it SDSS J110912.40+612346.7} ($z~=~0.0067$): Also known as UCG~06192 or MCG +10-16-069. BGH08 identified this object as a nearly identical type 2 version of the nearby narrow-line Seyfert 1 (NLS1) NGC~4395. Both objects have late-type spiral host galaxies and high-excitation optical spectra with narrow-line ratios consistent with low metallicities, with the exception that \eleven\ shows no evidence for broad-line emission. It also has the lowest stellar mass of any Seyfert 2 galaxy in the \citet{Kauffmann} SDSS AGN catalog, with log\ $(M_{*}/\msun)=8.07$ and a host galaxy luminosity of $M_g = -16.8$ mag. However, if we estimate the stellar mass using the prescription of \citet{Bell03} as above, and the SDSS catalogue Petrosian magnitudes ($u = 16.66, r = 15.21$ and $z=15.59$), we find $\mathrm{log}\ (M_\star/\msun) \approx 8.69$. BGH08 were unable to measure \sigmastar\ for this galaxy, but they demonstrate that the well-established correlation between \sigmastar\ and [\ion{O}{3}] linewidth for Seyfert 2 galaxies \citep{NW96} holds for galaxies with stellar velocity dispersions as low as $\sigmastar \sim 40-80$ \kms.  From the measured $\mathrm{FWHM}$([\ion{O}{3}])\ $= 66 \pm 1$ \kms, this correlation suggests a stellar velocity dispersion of $\sigmastar \sim \mathrm{FWHM}$([\ion{O}{3}])$/2.35 = 28$ \kms. BGH08 also obtained spectropolarimetry data for this object, finding a significant polarized continuum component, but no polarized emission-lines from a hidden broad-line region. In the same observation, the blue spectrum shows [\ion{Ne}{5}] $\lambda3426$ emission, giving additional evidence for AGN activity.

{\it SDSS J144012.70+024743.5} ($z~=~0.0297$): Also known as Tol 1437+030. This galaxy was also previously detected by the \rosat\ All-Sky Survey and a high-resolution Keck spectrum shows evidence for possible, but very weak broad \hal\ emission (BGH08). Its spectrum is also very similar to those of NGC~4395 and POX~52, with similar narrow-line ratios and detected high-ionization lines, but much weaker broad \hal\ emission. We again use the prescription of \citet{Bell03} and the SDSS catalogue Petrosian magnitudes ($u = 18.02, r = 16.40$ and $z=15.91$) to estimate a stellar mass of $\mathrm{log}\ (M_\star/\msun) \approx 9.85$ for this object. We note that the AGN light is relatively bright in this object and therefore our stellar mass estimate should reflect an upper limit to the actual stellar mass of the galaxy.

\section{Observations and Data Reduction}
Each object was observed using the EPIC instrument on \xmmn\ for $\sim 25$ ks. Due to the low signal-to-noise (S/N) of our observations, we focus our attention on the data taken with the EPIC-pn instrument, which has a higher quantum efficiency  than the EPIC-MOS instruments. Observation dates and actual exposure times (with soft proton flares excluded) for each object can be found in Table \ref{obsDate}. At most, $32.5\%$ of the exposure time was lost due to soft proton flaring in \one, $25.3\%$ was lost in \fourteen, $19.4\%$ was lost in \ten,  and no time was lost in \eleven. All data were reduced using the Science Analysis System (SAS, version 7.1.0) and XSPEC (version 12.4.0aa) following the guidelines of the SAS Cookbook and SAS ABC-Guide. Each object was extracted in a circular region with a radius of $30\arcsec$. Due to the proximity of all four objects to the edge of their respective chips, background regions free of sources were used with radii of $45\arcsec$, located on the same chip an average of $107\arcsec$ away from the source. Only events corresponding to patterns $0-4$ (single or double pixel events) were used for the pn event files of objects \ten\ and \fourteen. The event files for the other two objects, \one\ and \eleven, were limited to events corresponding to patterns $0-12$ (single, double, triple and quadruple pixel events) in order to maximize the S/N. In addition to pattern filtering, events were excluded that occurred next to the edges of the CCD or next to bad pixels, and all event files were restricted to an energy range of $0.3-10.0$ keV. The source count rates for all objects were low enough to neglect the effects due to pile-up.

\section{Results}
No counts above background are detected within the extraction region of \one. Assuming Possion statistics we estimate a 3$\sigma$ upper limit of $<~6.6$ counts from the background in the source extraction region, using the method of \citet{Gehrels}. The other three objects are each detected with a $>~3\sigma$ significance in the $0.5-10$ keV energy range, with the net counts in the range of $27-163$ for each detected object. 

\begin{figure}
\epsscale{1.0}
\plotone{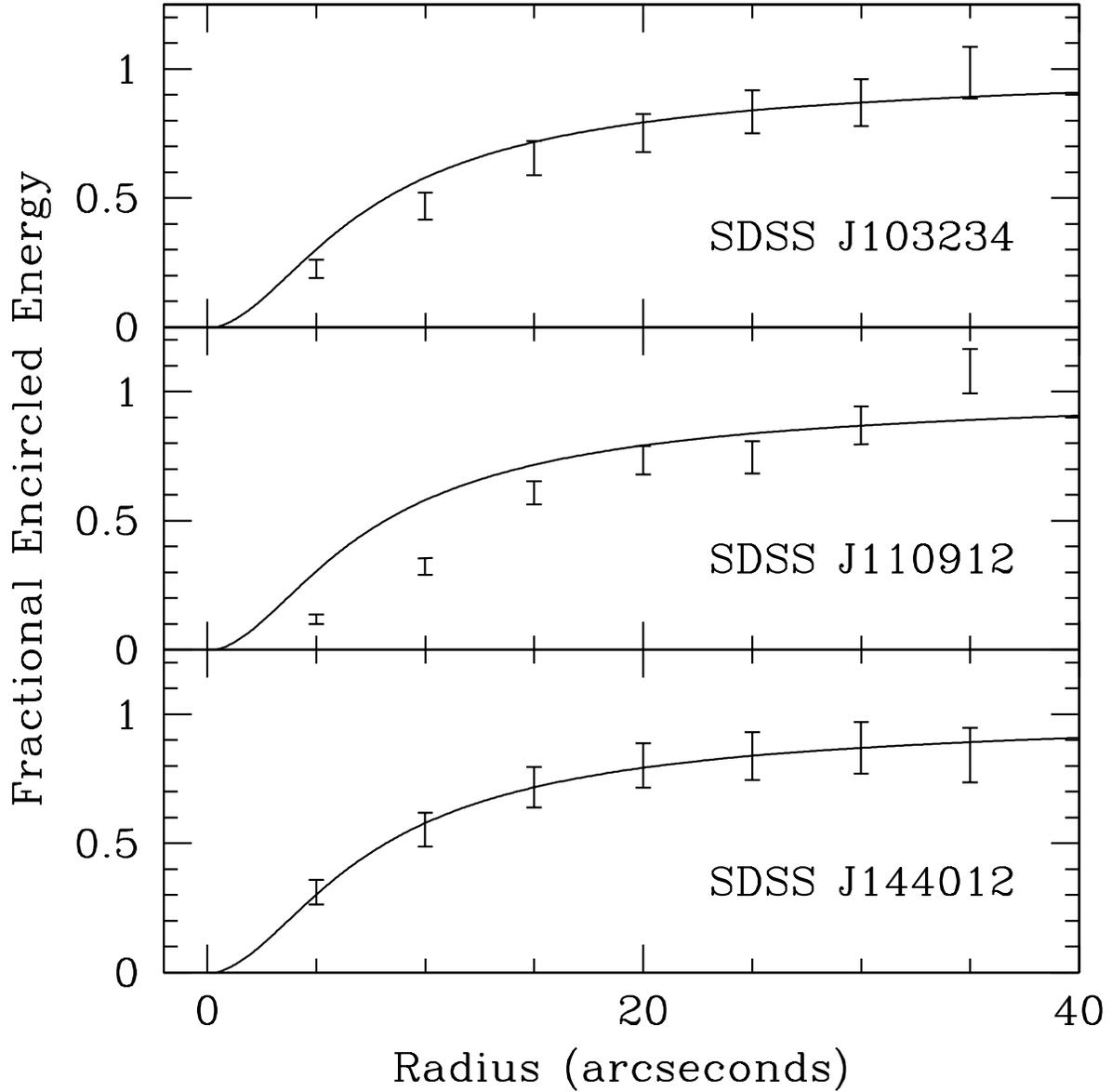}
\caption{Fraction of encircled energy as a function of radius. The solid line shows the encircled energy curve for the \xmmn\ PSF. The data points are normalized such that 87\% of the total energy is encircled at a radius of $R = 30\arcsec$.}
\label{encirc}
 \end{figure}

\subsection{Radial Profiles}
We use the SAS task {\it calview} to determine the EPIC-pn point spread function (PSF) at the time and CCD chip location of each of the three detected objects. For each object, we plot the encircled energy curve for the \xmmn\ PSF along with the fraction of encircled energy as a function of radius (Figure \ref{encirc}). The radius that encircles $50\%$ of the flux from a point source is $\sim 10\arcsec$, which translates to a physical diameter of $\sim 3$ kpc for \ten\ and \eleven, and $\sim 12$ kpc for \fourteen. This covers a considerable portion of the disk of \ten, more than half of the disk of \eleven, and the entirety of \fourteen. We see no evidence for extended emission outside of the physical extent of the PSF, except in object \eleven\ where we see a strong mismatch between the fractional encircled energy of the object and what is expected from a point source. The slope of the data is much steeper than that expected from the PSF, suggesting that we see only extended emission and not a PSF-dominated source. This may be due to high obscuration of the central engine and a few X-ray binaries within the source region contaminating our detection of the central engine.

\begin{deluxetable}{lccccccccc}
\rotate
\tabletypesize{\footnotesize}
\tablewidth{9.5in}
\tablecaption{X-ray Parameters}
\tablehead{
\colhead{Galaxy} & \colhead{X-ray} & \colhead{Net} & \colhead{Count Rate ($s^{-1}$)} 
& \colhead{$C_S~(s^{-1}$)} & \colhead{$C_H~(s^{-1}$)} & \colhead{HR} 
& \colhead{$\Gamma_\mathrm{HR}$} &\colhead{Flux ($\mathrm{erg~cm^{-2}~s^{-1}}$)} 
& \colhead{$L_\mathrm{X}~(\mathrm{erg~s^{-1}})$} \\
\colhead{} & \colhead{offset (\arcsec)} & \colhead{Counts} & \colhead{} & \colhead{} & \colhead{} & \colhead{} & \colhead{} & \colhead{} & \colhead{}
}
\startdata
SDSS J011905 & \nodata	& $<$ 6.6 & $<$ 0.0004          & \nodata             & \nodata             & \nodata              & \nodata      & $< 5.9 \times 10 ^{-15}$        & $< 1.3 \times 10 ^{40}$ \\
SDSS J103234 & 2.28	& 140.8             & $0.0065 \pm 0.0008$ & $0.0032 \pm 0.0005$ & $0.0033 \pm 0.0005$ & $\phn0.01 \pm 0.06$ & $1.1 \pm 0.1$ & $(3.4 \pm 0.7) \times 10^{-14}$ & $(2.5 \pm 0.5) \times 10^{39}$ \\
SDSS J110912 & 1.83	& \phn26.7          & $0.0012 \pm 0.0006$ & $0.0007 \pm 0.0004$ & $0.0005 \pm 0.0005$ & $-0.13 \pm 0.16$    & $1.4 \pm 0.2$ & $(4.7 \pm 2.6) \times 10^{-15}$ & $(4.6 \pm 2.5) \times 10^{38}$ \\
SDSS J144012 & 0.64	& 163.2             & $0.0082 \pm 0.0009$ & $0.0059 \pm 0.0007$ & $0.0023 \pm 0.0006$ & $-0.43 \pm 0.05$    & $1.8 \pm 0.1$ & $(1.7 \pm 0.4) \times 10^{-14}$ & $(3.4 \pm 0.9) \times 10^{40}$ \\
\enddata
\tablecomments{X-ray offset is the positional difference between the optical and X-ray positions. Total counts is the number of counts in the $0.5-10.0$ keV energy range, $C_S$ is the count rate from $0.5-2.0$ keV and $C_H$ is the count rate from $2.0-10.0$ keV. Flux and $L_{X}$ are based on the $2-10$ keV energy range and are inferred from a power law with photon index $\Gamma_{\rm HR}$ and Galactic absorption.}
\label{counts}
\end{deluxetable}

\subsection{Hardness Ratios}
Separating the events by energy, we investigate the hard ($C_H,~2.0-10.0$ keV) and soft ($C_S,~0.5-2.0$ keV) count rates using the 
hardness ratio (HR $= [C_H - C_S]/[C_H + C_S]$). \fourteen\ showed the softest spectrum in the 
sample, with HR $= -0.43 \pm 0.05$. The other two objects, \ten\ and \eleven, each had similar hard and soft count rates, resulting in HR $= 0.01 \pm 0.06$ and HR $= -0.13 \pm 0.16$, respectively.

We use the \xmmn\ response matrices, auxiliary response file (ARF) and redistribution matrix file
(RMF), to create model spectra in order to calculate a photon index ($\Gamma_{\rm HR}$) from the
HR following the method of \citet{Gall05}. These models assume the neutral absorber is set to 
the Galactic value in the direction of the object and the AGN is described by a simple power law. We caution that the intrinsic slope of the power-law continuum may not be well described by this simplistic model if the spectrum contains more complex components, such as a high level of absorption or a soft excess due to a thermal component. We include this analysis as a simple indicator of spectral slope, particularly for those objects where detailed spectral modeling is not feasible due to low S/N. \fourteen\ has a photon index ($\Gamma_{\rm HR} = 1.8 \pm 0.1$) that is similar to many 
Seyfert 1 galaxies, including POX 52, which had $\Gamma_{\rm HR} = 1.7$ when observed in an unobscured state with \chandra\ \citep{Thornton}. X-ray surveys of unobscured AGNs have found average power-law slopes of $\Gamma =1.9$ \citep{Nandra97, Nandra05}. \ten\ shows a much harder 
photon index ($\Gamma_{\rm HR} = 1.1 \pm 0.1$), similar to that seen in NGC~4395 
\citep{Moran05}, and \eleven\ has a photon index  of $\Gamma_{\rm HR} = 1.4 \pm 0.2$.
All count rates, HR, and $\Gamma_{\rm HR}$ can be found in Table \ref{counts}.

\subsection{Flux Estimates}
We estimate the $2-10$ keV flux from the same model spectra used to derive the photon indices from the HR. Again, these model spectra assume a neutral absorber set to the Galactic value and a power law with a slope derived as above and do not account for any additional components present in the spectra. The X-ray luminosities (\lX) are derived from these fluxes accounting for the distance of the object and Galactic absorption corrections are negligible. These flux estimates are consistent with those derived using the energy conversion factors calibrated by \citet{Hasinger} and found in the {\it XMM-Newton User's Handbook}. Individual flux and \lX\ estimates can be seen in Table \ref{counts}. 

\begin{figure}
\epsscale{1.0}
\plotone{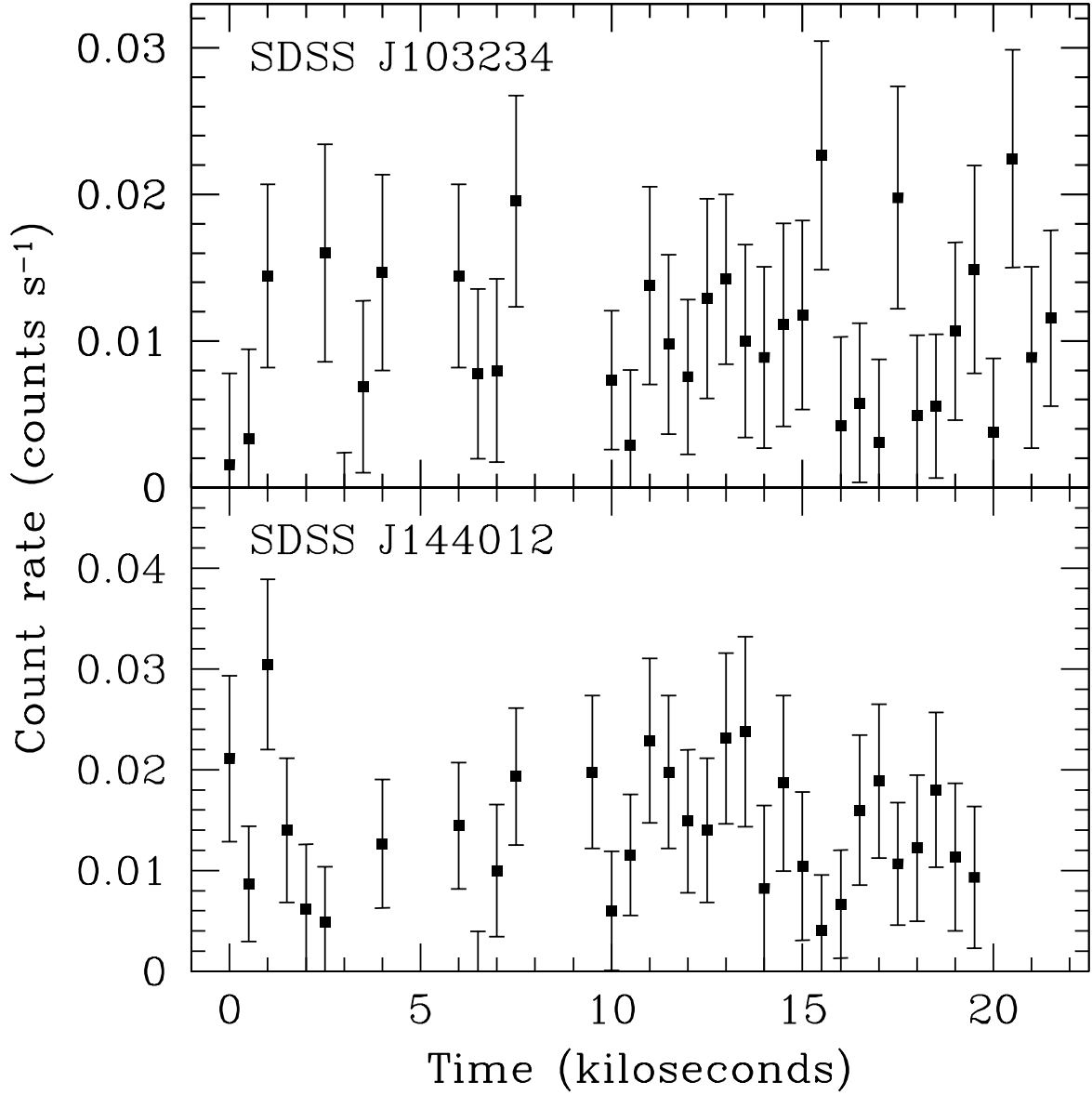}
\caption{Light curves of \ten\ ({\it top panel}) and \fourteen\ ({\it bottom panel}) binned by 
500 s each.}
\label{LightCurve}
\end{figure}

\subsection{Light Curves}
The two brightest objects in the sample, \ten\ and \fourteen, have high enough count rates to search 
for temporal variations. Using 500 s time bins, we create light curves 
(Figure \ref{LightCurve}) for these two objects, excluding any background flares present during 
the observations. We attempt to quantify any variability in these sources using the normalized 
excess variance ($\sigma^{2}$) of \citet{Nandra}:
\begin{equation}
\sigma^2_\mathrm{nxs} = \frac{1}{N \mu^2} \sum_{i=1}^{N} [(X_i - \mu)^2 - \sigma_i^2],
\label{excessvariance}
\end{equation}
where $X_i$ denotes the count rate of the $i$-th point in the light curve, $\sigma_i$ is its 
uncertainty, $\mu$ is the mean of the $X_i$ values over the entire light curve, and $N$ is the 
number of points in the light curve. The excess variance using 500 s time bins for \ten\ and \fourteen\ is $\sigma_{nxs}^{2}~=~-0.09~\pm~0.76$ and $\sigma_{nxs}^{2}~=~-0.56~\pm~1.2$, respectively. In order to increase the S/N in each data point, we enlarge the time bins to 1000 s and calculate an excess variance of 
$\sigma_{nxs}^{2}~=~1.3~\pm~13$ and $\sigma_{nxs}^{2}~=~0.08~\pm~1.5$ for \ten\ and \fourteen, 
respectively. In each case, $\sigma_{nxs}^{2}$ is consistent with zero, meaning that there is no evidence for intrinsic source variability. 

Previous \rosat\ detections of \ten\ and \fourteen\ as part of the \rosat\ All Sky Survey suggest a decrease in flux over a $\sim 10$ year time period. Assuming a photon index in the range of $\Gamma = 1-2$ and no absorption or thermal components, the estimated luminosity in the $0.1-2.4$ keV band from \rosat\ is $L_{X} = (0.6 - 2.8) \times 10^{40}$ \ergs\ and $L_{X} = (1.8 - 8.7) \times 10^{41}$ \ergs\ for \ten\ and \fourteen, respectively. \xmmn\ images of \ten\ and \fourteen\ show no other objects with similar or larger fluxes within the large (96\arcsec) PSF of the \rosat\ All Sky Survey, so contamination of the \rosat\ luminosities is unlikely. This is up to an order of magnitude larger than the luminosities derived in this work and could be larger if a substantial amount of absorbing material were present at the time the \rosat\ observations were taken. Variations of this magnitude have been observed before and are typically explained by variations in the absorbing material, especially at such soft energies. The Seyfert 2 galaxy NGC 4388 was observed to have a factor of $\sim 10$ increase in flux due to an order of magnitude decrease in the absorbing column density, typically at $N_{\rm H} \approx\ 3~\times~10^{23}~\mathrm{cm^{-2}}$ \citep{Elvis04}, over the course of a year. Similarly, NGC 4358 has shown order of magnitude variations in the $0.5 - 2$ keV luminosity due to $\sim 25\%$ variations in the absorbing column density over a one month time period \citep{Fruscione05}. This object has also seen factors of $2-3$ variations in absorbed flux due to continuum variability and not changes in the obscuring material. Therefore, it is reasonable to expect that the change in flux for both \ten\ and \fourteen\ are due to either continuum or absorbing column density variability or a combination of both.

\subsection{Spectral Fitting}
\ten\ and \fourteen\ are the only objects in the sample with high enough S/N for spectral
fitting, albeit over a limited range of energy. In both objects, the source spectrum becomes indistinguishable from that of the background at the high-energy end, due to low source counts. We choose to minimize the Cash $C$ statistic \citep{Cash79} in order to optimize the spectral fits instead of the often used $\chi^{2}$ statistic because it does not require a minimum number of counts per bin and the results are independent of the bin size used (for further discussion, see Cash 1979). Therefore, the $C$ statistic is more reliable than the $\chi^{2}$ statistic when fitting low S/N spectra. 

\begin{figure}
\begin{center}
\epsscale{1.1}
\plotone{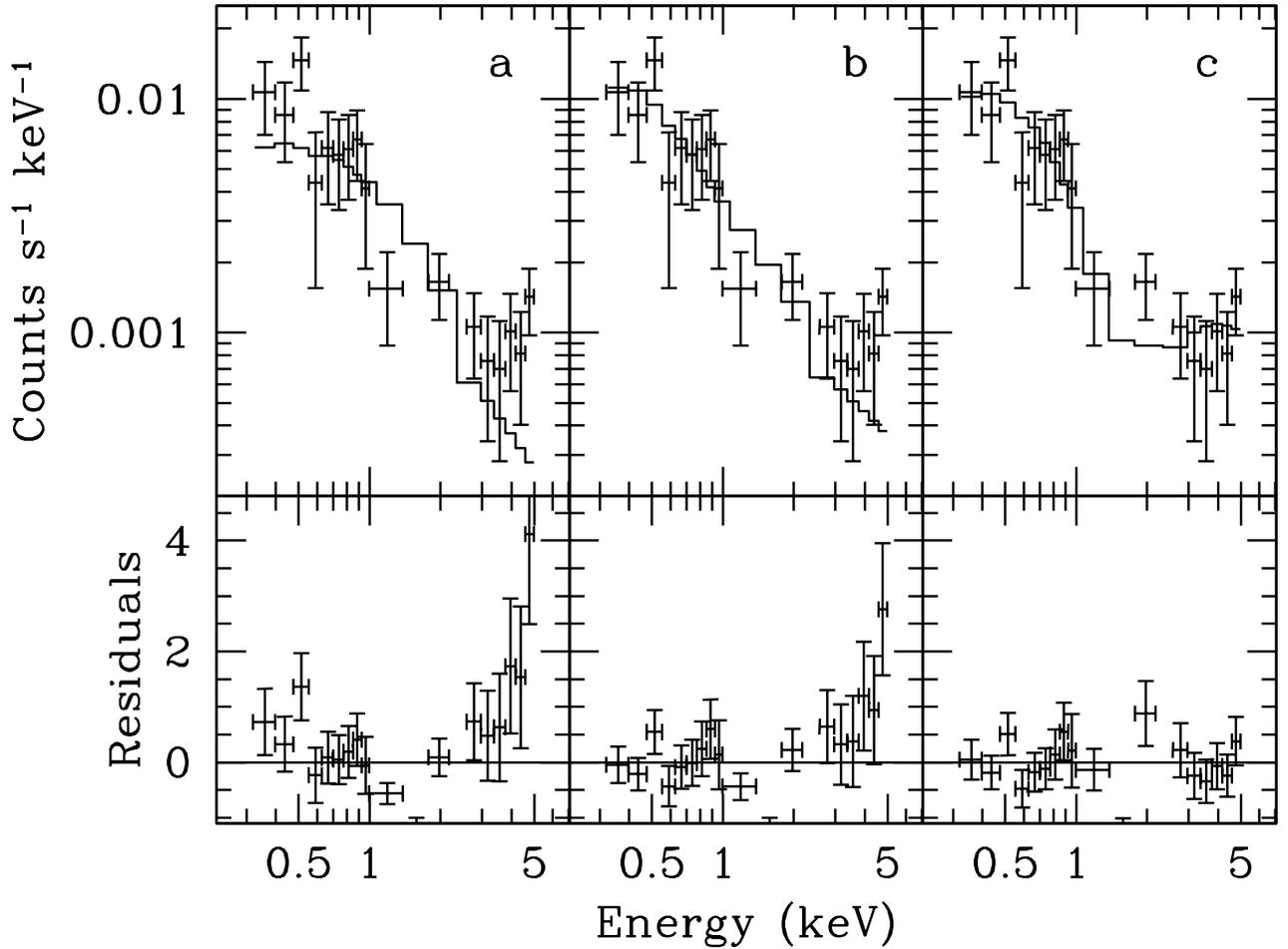}
\end{center}
\caption{The spectrum of \ten\ from $0.3-5.0$ keV modeled ({\it black line}) with ({\it a}) 
an absorbed power law with absorption set at the Galactic value, ({\it b}) an absorbed (Galactic 
value) power law (photon index set to $\Gamma = 1.1$) with a thermal disk blackbody, and 
({\it c}) an absorbed (Galactic value) power law ($\Gamma = 1.1$) with a thermal disk blackbody 
and a partial covering absorption component. Residuals are calculated as 
Residual = (Data - Model)/Model.}
\label{1032spec}
\end{figure}

\subsubsection{\ten}
Due to low S/N, \ten\ could only be fit over the $0.3-5.0$ keV energy range and therefore the data were rebinned to 15 channels per bin at energies $< 1.0$ keV and 80 channels per bin for energies $> 1.0$ keV in order to increase the S/N in each bin. At energies above $\sim 5$ keV, the number of source counts in a given bin were comparable to the background level and therefore, no useful spectral information could be extracted. Because of the degeneracies involved in fitting such a small spectral region with multiple components, we first fit a simple absorbed power law. We first allowed the absorbing column density to vary freely, but this produced an absorbing column density of zero. Therefore, we fixed the absorbing column density to the minimum value possible, the Galactic value of $N_{\rm H} = 1.18 \times 10^{20}~\mathrm{cm ^{-2}}$ for all following model fits. The best fit ($C = 54.32$ using 18 PHA bins and 16 dof) was a $\Gamma = 1.6 \pm 0.5$ power law that shows strong, systematic residuals at the soft and hard-energy ends of the spectrum (see Figure \ref{1032spec}). X-ray studies of low-mass type 1 Seyfert galaxies often show a soft excess, presumably due to a thermal component, with similar temperatures and strengths to those seen in narrow-line Seyfert 1 galaxies \citep{GH07a, Thornton, Mini, Desroches}. Modifying the model to include a thermal disk blackbody and allowing all parameters to vary freely (except for the column density for reasons discussed above) resulted in an unrealistic photon index of $\Gamma = -0.7^{+0.7}_{-1.0}$. Using the same absorbed power law with a thermal blackbody model, we tested setting the power-law slope to $\Gamma = 1.0$, increasing this slope by $0.5$ increments in subsequent model fits until $\Gamma = 3.0$, while allowing the blackbody temperature to vary freely. This allowed us to test more complex models over our limited energy range, while still keeping the number of free parameters to a minimum. Using this technique, we found the model fit with the photon index set at $\Gamma = 1.0$ produced the best result and therefore, we fix the photon index to $\Gamma = 1.1$, the value derived from the HR, for all further model fits. Fixing the photon index to this value in the absorbed power law and thermal blackbody model improves slightly ($C = 37.63$ using 18 PHA bins and 15 dof) with a blackbody temperature of $kT = 0.14 \pm 0.05$ keV, but still slightly underpredicts the spectrum at energies of $>~2.0$ keV. 

Finally, we note that this spectrum flattens out at energies $> 1.0$ keV, similar to the flattening seen at energies between $1-5$ keV in the \xmmn\ spectrum of POX~52 when it was observed to be in a partially-covered state \citep{Thornton}. With this in mind, we add a partial-covering component to our model and keeping the Galactic absorber fixed, we again test the incremental photon index values, finding similar results as before. We therefore fixed the photon index to value derived from the HR, $\Gamma = 1.1$ and find the best fit ($C~=~17.93$ using 18 PHA bins and 13 dof) from this model includes a $kT~=~0.20$ keV blackbody and an additional absorbing column density of $N_{\rm H}~=~4.3~\times~10^{22}~\mathrm{cm ^{-2}}$ covering $95\%$ of the X-ray-emitting region. This model is more complex than the previous models tested, but better fits the flattened region between energies $1-5$ keV. If this model accurately describes the observed emission, we would expect to see the spectrum of \ten\ decrease at energies $> 5$ keV and follow the shape of the power-law component. Additional observations with a higher S/N are needed to confirm this prediction, but based on the data currently available, we select the partially-covered absorbed power law as our best-fit model. We also note that we modeled the spectrum using multiple background regions surrounding the source and saw no significant changes that might affect our conclusions as to the best-fit model.

\begin{figure}
\begin{center}
\epsscale{1.1}
\plotone{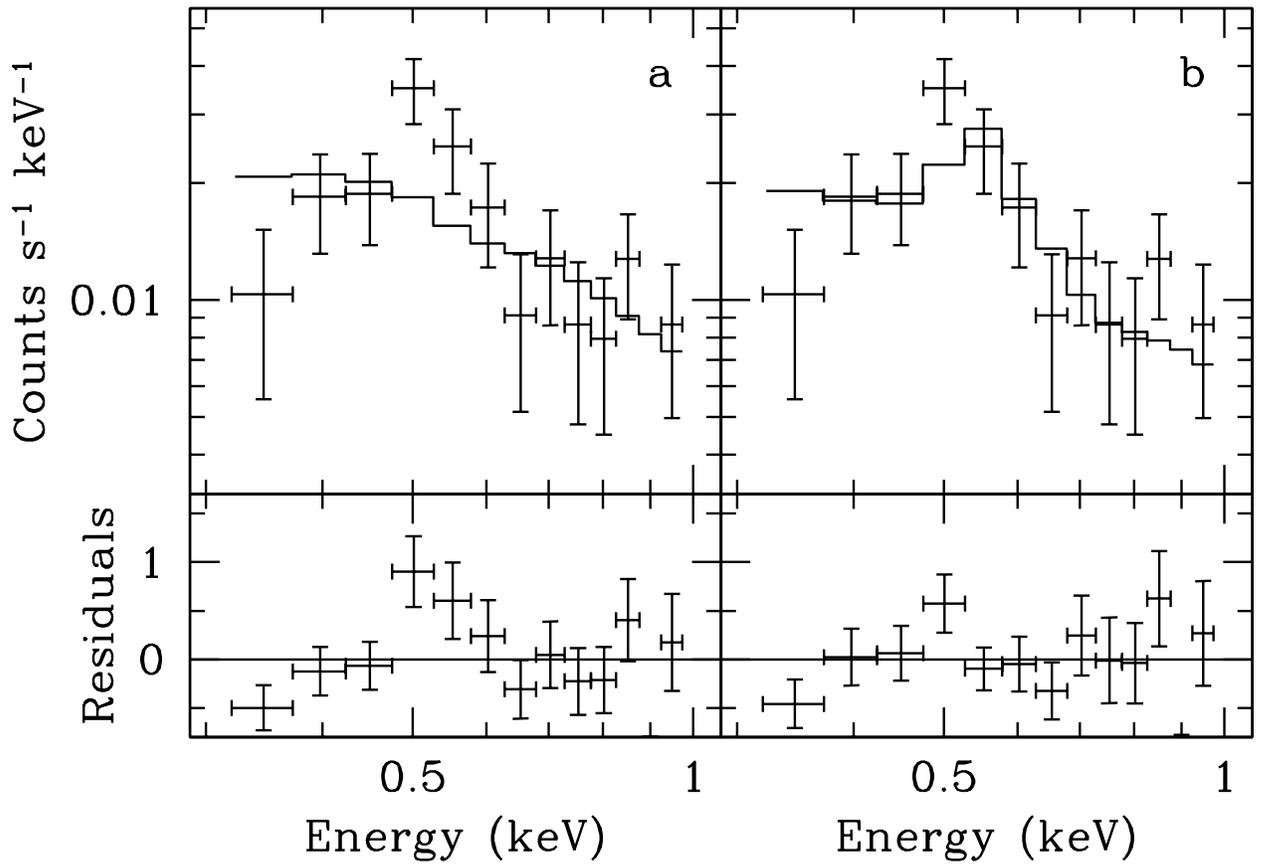}
\end{center}
\caption{The spectrum of \fourteen\ from $0.3-1.0$ keV modeled ({\it black line}) with 
({\it a}) an absorbed (set at the Galactic value) power law, and ({\it b}) an absorbed (Galactic
value) power law with a Raymond-Smith plasma. Residuals are calculated as 
Residual = (Data - Model)/Model.}
\label{1440spec}
\end{figure}

\subsubsection{\fourteen}
The spectrum of \fourteen\ extends from $0.3-1.0$ keV, above which the S/N is too low for spectral fitting, and was rebinned to 10 channels per bin. We start with a simple absorbed power-law model with the absorption fixed to the Galactic value ($N_{\rm H}~=~2.90~\times~10^{20}~\mathrm{cm^{-2}}$), with the understanding that the best-fit power law may not accurately describe the intrinsic underlying continuum over the full $0.5-10.0$ keV range. The left panel of Figure \ref{1440spec} shows the spectrum of \fourteen\ over-plotted with the best-fit $\Gamma~=~2.7~\pm~0.5$ power law ($C~=~26.60$ using 13 PHA bins and 11 degrees of freedom, dof). This power law is steeper than what is commonly seen in Seyfert 1 galaxies \citep{Nandra97} and does not fit energies below $0.6$ keV well. We also tested allowing the column density to vary freely, but this did not improve the fit quality and so we choose the Galactic value for simplicity. Therefore, the simple absorbed power law is a poor model for describing the X-ray spectrum of this source. 

In order to better fit the soft-energy end of the spectrum, we added a thermal component to the absorbed power law. We first added a blackbody component to the absorbed power law, allowing the thermal temperature of the blackbody and the photon index of the power law to vary freely, while holding the column density fixed to the Galactic value, for the reasons discussed above. This model was unable to fit the peak in the spectrum, seen at $E \sim 0.5$ keV, despite the range of parameter values tested, resulting in typical Cash statistics of $C > 23$.

Next, we tested replacing the blackbody component with a hot, diffuse plasma model, using either the Raymond-Smith plasma \citep{Raymond} or the MEKAL plasma \citep{Mewe85, Mewe86, Kaastra92, Liedahl92} models. We tested both of these models individually, fixing the redshift parameter to $z=0.0297$ and the abundance parameter to the solar value. We again tested allowing the column density to vary freely, but found the best results were achieved when the absorbing column density was fixed to the Galactic value. All other parameters were allowed to vary freely in each component. Using the Raymond-Smith plasma component produced a best fit ($C~=~18.73$ using 13 PHA bins and 9 dof; Figure \ref{1440spec}) with a power-law photon index of $\Gamma = 1.5^{+0.7}_{-3.0}$ and a plasma temperature of $kT = 0.13^{+0.05}_{-0.02}$ keV. This temperature is within the range of typical values of $kT \sim 0.1 - 0.2$ keV seen in narrow-line Seyfert 1 galaxies and PG quasars \citep{Leighly, Piconcelli}. The model using the MEKAL plasma component resulted in a very similar best fit ($C~=~17.94$ using 13 PHA bins and 8 dof), with a MEKAL plasma temperature of $kT = 0.14$ keV and a corresponding photon index of $\Gamma = 1.0^{+1.4}_{-1.0}$. We note that the photon indices of both of these models are not well constrained with lower bounds of $\Gamma \leq 0$. This is most likely due to the very small energy range over which we are attempting to fit these models. Both model fits are nearly indistinguishable from each other, except that the MEKAL model uses one additional free parameter. Therefore, we select the model containing the Raymond-Smith plasma as our best-fit model for simplicity.

\begin{figure}
\epsscale{1.0}
\plotone{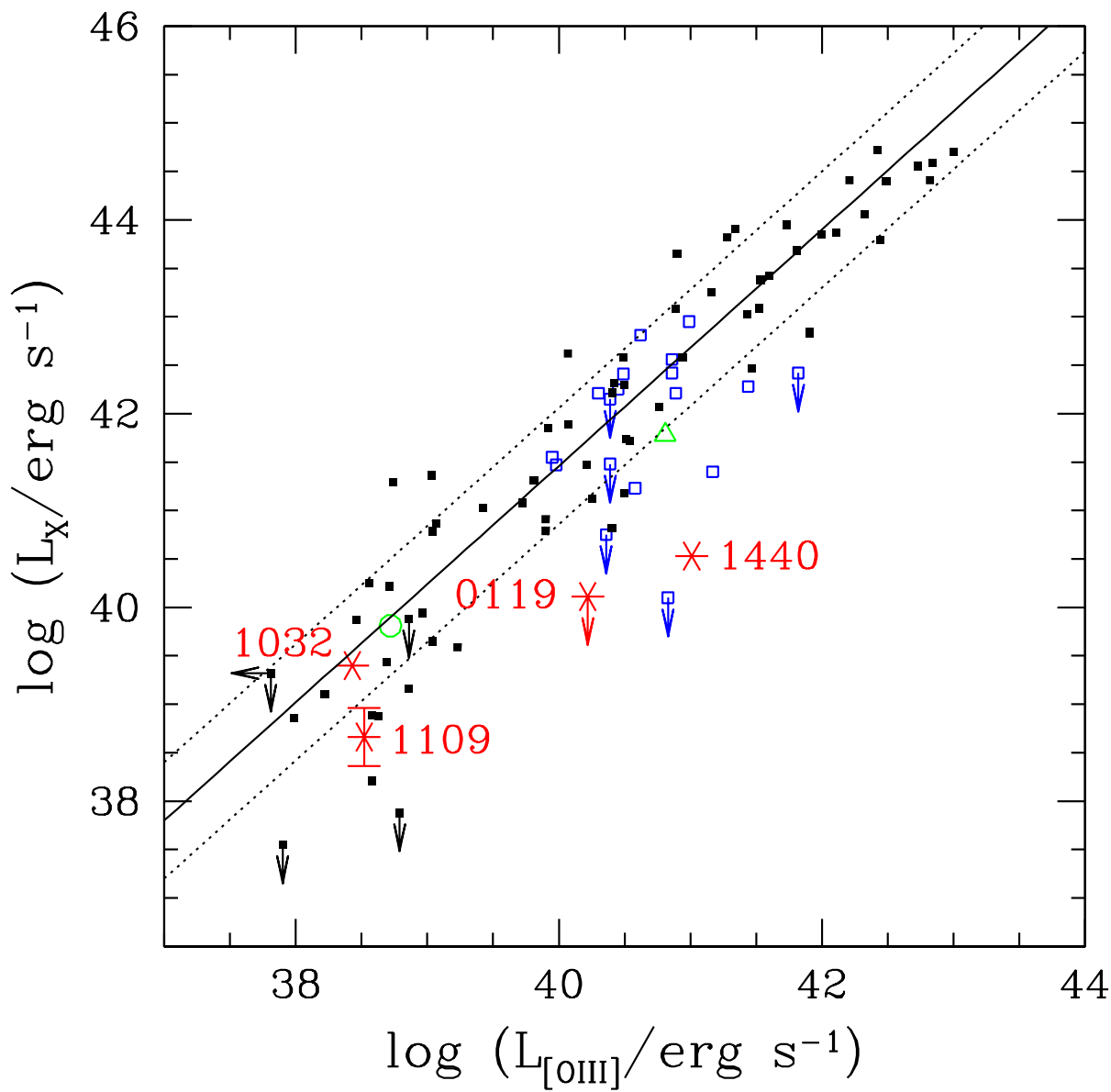}
\caption{Plot of \loiii\ vs ${\rm L_{2-10~keV}}$ with our sample ({\it red asterisks}) along with the objects ({\it black, filled squares}) from \citet{Panessa}, which includes quasars and Seyfert 1 and 2 galaxies, all corrected for X-ray absorption. The open symbols represent NGC~4395 ({\it green circle}; Panessa \etal\ 2006, corrected to the updated distance of 4.3 Mpc; Thim \etal\ 2004), POX~52 ({\it green triangle}; Barth \etal\ 2004, Thornton \etal\ 2008) and low-mass Seyfert 1 objects ({\it blue squares}; Greene \& Ho 2004, Desroches \etal\ 2009).
The solid line represents the least-squares fit derived by \citet{Panessa} and the two dotted lines represent the 1$\sigma$ scatter of the Panessa \etal\ sample about the fit. Unless otherwise displayed, error bars for our sample are smaller than the plot symbol.}
\label{LoiiiLx}
\end{figure}

\begin{deluxetable}{lccccccc}
\rotate
\tabletypesize{\footnotesize}
\tablewidth{8.4in}
\tablecaption{Derived Parameters}
\tablehead{
\colhead{Galaxy} & \colhead{\sigmastar~(\kms)} & \colhead{\mbh~(\msun)} & \colhead{\ledd~$(\mathrm{erg~s^{-1}})$} & \colhead{\lbol([OIII])~$(\mathrm{erg~s^{-1}})$} & \colhead{\lbol([OIII])/\ledd} & \colhead{\lbol(X-ray)~$(\mathrm{erg~s^{-1}})$} & \colhead{\lbol(X-ray)/\ledd} }
\startdata
SDSS J011905 & $39 \pm 8$ & $1.1 \times 10^{5}$ & $1.4 \times 10^{43}$ & $5.8 \times 10^{43}$ & \phn4.2      & $< 2.7 \times 10^{41}$         & $< 0.02$ \\
SDSS J103234 & $43 \pm 4$ & $1.6 \times 10^{5}$ & $2.0 \times 10^{43}$ & $9.3 \times 10^{41}$ & \phn\phn0.05 & \phn\phn$5.0 \times 10^{40}$   & \phn\phn\phn0.003  \\
SDSS J110912 & \phn$28 \pm 1$\footnotemark & $5.0 \times 10^{4}$ & $6.3 \times 10^{42}$ & $1.1 \times 10^{42}$ & \phn0.2      & \phn\phn$9.2 \times 10^{39}$   & \phn\phn\phn0.001  \\
SDSS J144012 & $45 \pm 4$ & $2.0 \times 10^{5}$ & $2.5 \times 10^{43}$ & $3.5 \times 10^{44}$ & 14.1      & \phn\phn$6.8 \times 10^{41}$   & \phn\phn0.03       \\
\enddata
\tablecomments{Stellar velocity dispersion measurements are from BGH08, except for \eleven, 
which is calculated from $\sigmastar~=~\mathrm{FWHM([OIII])}/2.35$. \lbol([OIII]) was estimated from
the [\ion{O}{3}] luminosity using the bolometric correction of $\lbol/L_\mathrm{[OIII]}~=~3500$.
\lbol(X-ray) was estimated using the bolometric correction $\lbol/L_\mathrm{X}~=~\kappa$, where
$\kappa~=~20$ \citep{VF07}.}
\footnotetext{This velocity dispersion was estimated based on the FWHM of the [\ion{O}{3}] line.}
\label{Lboltable}
\end{deluxetable}

\section{Discussion}
\subsection{\loiii-\lX\ Correlation}
The relationship between the luminosity of the [\ion{O}{3}] emission line at $5007$ \AA\ (\loiii) and the 
$2-10$ keV luminosity has been studied for a range of objects, including type 1 and 2 Seyfert galaxies and quasars, to determine if \loiii/\lX\ is similar among all Seyfert galaxies or whether the relationship has any additional dependence on properties such as accretion rate, luminosity, and black hole mass. \citet{Kraemer} investigated the range of X-ray and [\ion{O}{3}] luminosities in both broad-line and narrow-line Seyfert 1 galaxies, finding little difference between the two populations. \citet{Heckman05} followed this study with another, in which they included both Seyfert 1 and 2 galaxies, specifically investigating if this relationship extended to Seyfert 2 galaxies and whether or not \lX\ needed to be corrected for absorption. They found that if the \lX\ of a Seyfert 2 galaxy was corrected for absorption, the correlation remained intact with minimal scatter added due to the uncertainties involved in the absorption correction. \citet{Panessa} included a wider variety of objects to their sample in order to include a larger luminosity range than previous used, including objects with \lX $\sim 10^{37-38}$ \ergs, and found that the correlation remained approximately the same. An important outcome of this is that the optical [\ion{O}{3}] luminosity can be used as a tracer of the intrinsic X-ray luminosity, and therefore used to estimate the amount of absorption seen in the X-ray. We note that both the \citet{Heckman05} and \citet{Panessa} samples include a selection of radio-loud and radio-quiet AGN, while our four objects are defined as radio-quiet using the standard \citet{Kellermann} definition and the [\ion{O}{3}] luminosity to infer an optical luminosity. 

We plot the \loiii\ measurements for our four objects against our estimates of \lX\ along with the \loiii-\lX\ relation derived by \citet{Panessa} in Figure \ref{LoiiiLx}. All four objects are X-ray weak with respect to the relation, although \ten\ is within the $1\sigma$ scatter of the relationship, which is 0.6 dex in \lX\ at fixed \loiii. \eleven\ is $\sim 0.9$ dex below the relation and \fourteen\ and the upper limit of \one\ are both $>2\sigma$ outliers, falling $\sim 2.2$ dex and $1.5$ dex below the relation, respectively. 

We use the \loiii-\lX\ correlation from \citet{Panessa} and calculate the expected $2 - 10$~keV luminosity from the observed \loiii\ values. The [\ion{O}{3}] luminosities are determined from the SDSS spectra and corrected for Galactic extinction \citep[BGH08]{Kauffmann}. Comparing these values to the \lX\ values measured from the \xmmn\ data, we can attempt to quantify the level of intrinsic absorption within each galaxy. The observed \lX\ of \ten\ is consistent with the \lX\ derived from \loiii, so no additional absorption is needed to reconcile the two \lX\ values. \eleven\ needs an absorbing column density of $N_{\rm H} = 8.8 \times 10^{21}~\mathrm{cm^{-2}}$ in order to bring the observed \lX\ to the value suggested by its [\ion{O}{3}] luminosity. Due to the low \lX\ of the upper limit of \one, a considerable amount of absorption, $N_{\rm H} > 1.5 \times 10^{22}~\mathrm{cm^{-2}}$, is needed to bring the observed and predicted \lX\ values into agreement. However, without a proper detection of \one, the magnitude of the intrinsic absorbing column density will remain unknown. \fourteen\ also falls below the \loiii-\lX\ relation, suggesting that there is a considerable amount of absorption in this object. Using our measured \lX, we estimate the absorbing column density needed to account for its displacement from the \loiii-\lX\ correlation to be $N_{\rm H} = 2.3 \times 10^{22}~\mathrm{cm^{-2}}$.

It is unclear whether an absorbing column density of $N_{\rm H} \sim 10^{22}~\mathrm{cm^{-2}}$ is large enough to completely obscure all of the broad-line emission from a source, or even how the X-ray absorbing column and optical extinction are related. X-ray and optical surveys of AGN find $10-20\%$ of objects show the properties of one AGN type in the optical and another in X-ray, {\it e.g.} a narrow-line AGN with no absorption in the X-ray \citep{Perola, Silverman, Tozzi}. Among well-studied bright Seyfert galaxies, there are examples of objects with substantial broad-line emission in the optical, but with moderate levels of absorption observed in the X-ray. NGC 3227 is one of these objects with obvious broad-line emission and an X-ray absorbing column density of $N_{\rm H} \approx 6.5 \times 10^{22}~\mathrm{cm^{-2}}$ (Gondoin \etal\ 2003, see Piconcelli \etal\ 2006 and Jim{\'e}nez-Bail{\'o}n \etal\ 2008 for further examples). NGC 3227 and galaxies with similar optical line ratios are typically classified as intermediate Seyferts, due to the relative flux of the broad and narrow components of the permitted lines. The difference in appearance between intermediate Seyferts and more typical Seyfert 1 galaxies is often attributed to some amount of obscuration. Whether the level of X-ray absorption seen in galaxies like NGC 3227 is related to the amount of optical extinction in other objects, such as those in our sample, remains uncertain.

If we compare our measurements with other \loiii-\lX\ correlations, we find that the results can change substantially. For example, using the relationships derived by \citet{Heckman05} or \citet{Netzer}, we find that the expected X-ray luminosity calculated from the observed \loiii\ values are an order of magnitude lower than if the \citet{Panessa} relationship is used, which would suggest little to no absorption is present in any of our objects. However, both the \citet{Heckman05} and \citet{Netzer} samples contain objects more luminous than \lX $\sim 10^{41}$ \ergs, up to 1-3 orders of magnitude larger than observed in our objects. \citet{Netzer} also found a luminosity dependance in the \loiii/\lX\ ratio, which would explain different results for samples with substantially different ranges in luminosity. The \cite{Panessa} sample includes objects with the same range of X-ray and [\ion{O}{3}] luminosities as our sample, making it the more appropriate sample for comparisons.

\subsection{X-ray Binary Contamination}
X-ray observations of low-mass AGNs, including NGC~4395 \citep{Moran05}, POX 52 \citep{Thornton} and SDSS-selected objects \citep{GH07a, Desroches, Mini}, show that the type 1 population is predominantly low-luminosity, with typical luminosities of $\lX\ \approx 8 \times\ 10^{39} - 10^{43}$ \ergs. The X-ray luminosity is much lower for the population of type 2 AGNs in this sample ($\lX\ \approx 5 \times\ 10^{38} - 4 \times\ 10^{40}$ \ergs) and is close to the collective luminosity one would expect from a population of X-ray binaries in a host galaxy. This possible contamination is amplified by the fact that the \xmmn\ PSF covers a large fraction of each galaxy in our sample. We now consider how much of the observed X-ray flux might arise from non-nuclear sources in each of our host galaxies.

\citet{Gilfanov04} examined the relationship between the X-ray luminosity and stellar mass of old populations, and found a typical ratio of $\lX/M_\star\ = 8.3 \times\ 10^{28}$ \ergs\msun\per. Given the stellar mass range of our sample, $\mathrm{log}\ (M_\star/\msun) = 8 - 10$, it is unlikely that the observed X-ray fluxes are significantly contaminated by low-mass X-ray binaries associated with the old stellar population. However, high-mass X-ray binaries associated with recent star formation can result in a higher X-ray luminosity. \citet{Lehmer} examined the relationship between the $0.5-8$ keV X-ray luminosity of late-type, star-forming galaxies and their stellar mass. They found $\lX/M_\star\ = 1.6 \times\ 10^{30}$ \ergs\msun\per\ for star-forming galaxies with stellar masses of $\mathrm{log}\ (M_\star/\msun) = 9 - 10$.

Of the objects in our sample, the contamination by X-ray binaries is potentially most significant for \eleven. It has a late-type morphology and the bluest host galaxy colors in the BGH08 sample (with $g-r \approx 0.3$). Based on the Lehmer \etal\ results and the estimated host galaxy mass described in \S2, the predicated luminosity due to high-mass X-ray binaries is close to (and might even exceed) the observed X-ray luminosity, indicating that a large fraction of the observed X-ray flux might be non-nuclear. Additionally, the observed X-ray flux is more spatially extended than the PSF, as seen in the steep slope of the encircled energy curve for the object (Figure \ref{encirc}). However, without deeper and higher-resolution X-ray data, we cannot clearly determine the relative amounts of nuclear and non-nuclear emission. In the other late-type disk galaxy in our sample, \ten, the radial profile of the X-ray emission is more consistent with a point-like source. If the $\lX/M_\star$ ratio found by Lehmer \etal\ applies to this galaxy, then X-ray binaries could in principle account for the observed X-ray luminosity. However, the SDSS image shows that most of the recent star formation in this galaxy is located in the spiral arms at distances of $\gtrsim 10\arcsec$ from the nucleus. If the X-ray luminosity of this galaxy was dominated by high-mass X-ray binaries then it would be noticeably extended in the \xmmn\ image rather than compact, so it appears unlikely that high-mass X-ray binaries make a significant contribution to the observed X-ray flux. For \fourteen, the predicted luminosity due to high-mass X-ray binaries is a factor of 3 smaller than the observed X-ray luminosity. Although we can not rule out the possibility that some observed properties, such as $\Gamma$ and HR, might be affected by the presence of high-mass X-ray binaries, we conclude that the observed X-ray flux is most likely dominated by emission from the AGN.

\subsection{Bolometric Luminosity and Eddington Ratio}
We investigate the accretion power of these objects by deriving estimates of their Eddington ratios,
\lbol/\ledd. The \msigma\ relation allows us to estimate a black hole mass from the stellar
velocity dispersion (\sigmastar) of the host galaxy \citep{Gebhardt, FM}. BGH08 used their measurements of \sigmastar\ along with the \msigma\ relation of \citet{Tre02} to estimate a black hole mass for each object, which includes an additional offset in black hole mass seen in other low-mass Seyfert 1 galaxies \citep{Barth05}. This offset probably reflects a flattening of the \msigma\ relation at low masses \citep{Wyithe06, GH06}, but without a more detailed study of the \msigma\ relation in this mass range, we follow the example of \citet{Barth05} and assume a uniform offset.

A common estimator of bolometric luminosity in the optical is based on the [\ion{O}{3}] $\lambda5007$ line luminosity because it is assumed to be dominated by unobscured AGN emission regardless of AGN type, rather than by emission from star-forming regions. We first estimate the bolometric luminosity from \loiii\ for each object using the bolometric correction of \lbol/\loiii~$=~3500$ derived by \citet{Heckman} from a low-redshift sample of Seyfert 1 galaxies and quasars. 

We also use a broad-band $2-10$ keV X-ray bolometric correction from \citet{VF07, VF09}. Because a larger wavelength range is used to estimate \lbol, X-ray bolometric corrections tend to produce better estimates of the intrinsic \lbol\ than corrections based on narrow-band luminosities, assuming the unabsorbed X-ray luminosity is used. The $2-10$ keV energy range also probes the primary continuum emission from the central engine producing a better estimate of the bolometric luminosity of the object. \citet{VF09} find that highly accreting objects tend to require a larger X-ray bolometric correction than those accreting at a lower rate. For \lbol/\lX~$=~\kappa$, objects with an intrinsic \lbol/\ledd~$>~0.2$ should have a bolometric correction of $\kappa~\sim~110$, objects with $0.1 <$\lbol/\ledd$< 0.2$ should have $\kappa~\sim~45$, and for those objects with intrinsic \lbol/\ledd~$<~0.1$, $\kappa~\sim~20$. We find that with this two-tier bolometric correction, no objects in our sample are accreting at a high enough rate to warrant the higher bolometric correction. Therefore, we apply the X-ray bolometric correction of \lbol/\lX~$=~20$ to the unabsorbed X-ray luminosities of each of our objects to obtain an estimate of the bolometric luminosity. For each object, estimates of stellar velocity dispersion, black hole mass, Eddington luminosity and bolometric luminosity can be seen in Table \ref{Lboltable}. 

The Eddington ratios from \lbol(X-ray) and the absorption-corrected X-ray luminosities are systematically lower than those determined from the [\ion{O}{3}] luminosity by a factor of $\sim 30-400$. Underestimating the intrinsic absorption in these objects could explain a small portion of this discrepancy, but it is mostly due to a difference in the bolometric corrections themselves. \citet{Lamastra09} recently evaluated the \loiii\ bolometric correction of \citet{Heckman}, finding that even low levels of extinction in the narrow-line region cause bolometric luminosities to be systematically over-estimated. By correcting \loiii\ for this extinction and including a dependance on luminosity, they find the \loiii\ bolometric corrections to be over an order of magnitude lower than in \citet{Heckman}. The resulting bolometric luminosities more closely resemble the bolometric luminosities derived using X-ray bolometric corrections. This luminosity-dependent [\ion{O}{3}] bolometric correction may resolve the discrepency observed between our bolometric luminosities derived from \loiii\ and \lX. Regardless of the bolometric correction used, \fourteen\ has the highest Eddington ratio in the sample with \lbol(X-ray)/\ledd$~\approx~0.03$. A careful look of the spectral energy distributions of these objects using recently observed \spitzer\ IRS spectra along with existing multi-wavelength data should better determine the bolometric luminosities of these sources.

\subsection{\eleven}
\eleven\ is an object of particular interest, not only for the strong similarity between it and NGC 4395, but also because it has the lowest detected X-ray luminosity in the sample. The radial profile of \eleven\ shows possible evidence for extended emission and the X-ray luminosity of $\lX\ \sim 5 \times 10^{38}$ \ergs\ is so low that it could be contaminated by a collection of X-ray binaries in the host galaxy. On average, low-mass X-ray binaries have photon indices in the range of $\Gamma \approx 1.5-1.8$ (e.g. Matsumoto 1997; Irwin \etal\ 2003) and high-mass X-ray binaries have $\Gamma \approx 1-2$ (e.g. Sasaki \etal\ 2003), both of which are comparable to the measured spectral slope of \eleven\ ($\Gamma = 1.4 \pm 0.2$). X-ray variability is usually a better distinguishing parameter between X-ray binaries and AGNs since variability scales with black hole mass \citep{McHardy} and a collection of X-ray binaries would not vary coherently. Unfortunately, our observations contain too few counts to properly investigate the temporal variability of \eleven\ to attempt to determine the degree of contamination due to the stellar population.

If the X-ray emission detected is related to AGN activity, it is still unclear whether the central engine is obscured or if we are observing a weakly accreting AGN with no broad-line region. The estimated black hole mass for \eleven\ is very tentative, since BGH08 were unable to measure a stellar velocity dispersion for this galaxy, and the \msigma\ relation used to estimate the black hole mass is poorly constrained at these low masses. Given the strong morphological and spectral similarities between NGC 4395 and \eleven, it is conceivable that they also share comparable black hole masses. Then, the defining difference between these two objects would be the lack of broad-line emission in \eleven\ (BGH08). If \eleven\ falls within the normal scatter of the \loiii-\lX\ relation and therefore there is little internal absorption, one could speculate that \eleven\ represents a nearly identical ``true'' type 2 version of NGC~4395 with no broad-line region. There have been a few suggested cases of type 2 objects without a broad-line region (see Ghosh \etal\ 2007, Gliozzi \etal\ 2007, Bianchi \etal\ 2008), but \citet{BN} recently show that many of the candidate ``true'' type 2 objects can also be fit with more complex spectral models, including high amounts of absorption. 

Various models attempt to explain how an accreting black hole might fail to form a broad-line region, typically due to a combination of low AGN luminosity and black hole mass (for a review, see Ho 2008). \citet{Laor} suggested that since the radius of the BLR decreases as the luminosity of an AGN decreases, there might exist a luminosity threshold below which the BLR would cease to exist. For a black hole with $\mbh \sim 10^{5}$~\msun, this luminosity would be $L_\mathrm{min} = 6 \times 10^{35}$~\ergs\ and $L_\mathrm{min} = 6 \times 10^{37}$~\ergs\ for a $10^{6}$~\msun\ black hole, both of which are $2-6$ orders of magnitude below the bolometric luminosities of NGC 4395 and \eleven. If \eleven\ is indeed a "true" type 2 Seyfert, this scenario can not explain the lack of the broad-line region.

Alternatively, if the BLR is formed due to a radiation-driven wind flowing off of the accretion disk, as has been suggested by \citet{MC97}, then at very low luminosities the outflow may decrease to levels unable to sustain the BLR. Nicastro (2000, see also Nicastro \etal\ 2003) has explored a model in which there exists a threshold \lbol\ below which the radiation-driven wind is unable to produce the broad-line emission. This suggests that at lower \lbol, unobscured AGNs would be "true" type 2 objects. The calculations of \citet{Nicastro00} show that this threshold lies at $\lbol/\ledd\ \lesssim 10^{-3}$, below which the objects would be unable to produce the broad-line region. 

Using the X-ray results, we can examine where these low-mass AGNs lie relative to the predicted thresholds for BLR formation. We find that \ten\ and NGC~4395 both have Eddington ratios close to the hypothetical threshold for BLR formation in the Nicastro model (NGC 4395: $\lbol/\ledd\ \approx 10^{-3}$; Peterson \etal\ 2005). However, optical spectra show definite broad-line emission in NGC~4395 and weak broad-line emission in \ten\ (BGH08). \eleven\ has a very uncertain Eddington ratio of $\lbol/\ledd\ \approx 0.001$ which lies at the threshold for BLR formation in the Nicastro model. Therefore, there exists the possibility that \eleven\ accretes at a low enough rate to explain the lack of broad-line emission. However, this seems unlikely given that \eleven\ and NGC 4395 have such similar narrow emission-line luminosities, and presumably similar ionizing luminosities. 

Given that the observed \lX\ of \eleven\ is an order of magnitude lower than that of NGC~4395, but they have nearly identical [\ion{O}{3}] luminosities, we conclude that the most likely explanation is that the central engine of \eleven\ is obscured, although we are unable to accurately determine the amount of X-ray absorption in \eleven. If a substantial fraction of the observed \lX\ is due to X-ray binaries, then that increases the discrepancy between \loiii\ versus \lX\ and strengthens the case for an obscured nucleus. \citet{ES06} present a model in which the radiation-driven wind produces the obscuring material, such that objects with \lbol\ $\lesssim 10^{42}$ \ergs\ would be unable to produce the obscuring torus. If the observed properties of \eleven\ are indeed the result of obscuration, then this would indicate that the obscuring torus could still persist even at AGN luminosities more than an order of magnitude below the threshold suggested by Elitzur \& Shlosman.

\section{Summary and Conclusions}
We find all four objects in the sample to be X-ray faint with X-ray luminosities approximately an order of magnitude lower than those seen in the \citet{GH04} sample \citep{GH07a}, but only two objects, \one\ and \fourteen\ show evidence of moderate to substantial absorption with estimated column densities of $N_{\rm H} \sim 10^{22}~\mathrm{cm^{-2}}$. \ten\ shows little evidence of absorption which is consistent in the context of the unified model with the previous detection of broad emission lines from high-resolution optical spectra. It is unclear without further observations whether \eleven\ is truly lacking a broad-line region or if it is absorbed and the emission detected is due to a combination of X-ray binaries in the host galaxy and weak emission from the AGN, but given the low observed X-ray luminosity, we believe \eleven\ contains at least a moderate amount of absorption. 

Two objects had high enough S/N ratios for spectral fitting. \ten\ has a spectrum similar to the partially absorbed spectrum of POX~52 discussed by \citet{Thornton}, and is well fit with a high partial-covering fraction of 95\%, in the $0.3~-~5.0$ keV range probed. The spectrum of \fourteen\ was well fit over an energy range of $0.3-1.0$ keV with a power law combined with a thermal plasma component and Galactic absorption and there was no clear evidence from the spectral fit for the high absorption predicted from the \loiii-\lX\ correlation. 

By comparing type 1 and 2 objects with similar masses, we have the opportunity to further investigate the absorbing properties of low-mass AGNs and whether or not line-of-sight absorption can explain the presence or lack of broad emission lines, or if other processes, such as accretion rate, play a role. If high S/N observations are obtained, X-ray spectra are an excellent tool that can be used to quantify the amount of absorption in an object. IR spectroscopy is another important tool used to investigate absorption and reprocessing of the AGN continuum. We plan to investigate both low-mass Seyfert 1 and 2 objects using {\it Spitzer} spectra in order to better constrain the bolometric luminosities of these objects and further investigate the presence of an obscuring torus.

\acknowledgments The analysis of \xmmn\ data presented herein was supported by grant 
NNX06AF08G from NASA. This work was also supported by the National Science Foundation under grant AST-0548198. This research has made use of the NASA/ IPAC Infrared Science Archive, which is operated by the Jet Propulsion Laboratory, California Institute of Technology, under contract with the National Aeronautics and Space Administration.

\clearpage

\end{document}